\documentstyle{mn}

\def\bmit{\bf}
\def\dta{\delta}
\def\v{_{\theta}}
\def\ii{{\rm i}}
\def\d{{\rm d}}

\def\vr{{\bf r}}           %\def\vr{{\vec r}}         
\def\vn{{\bf n}}           %\def\vn{{\vec n}}         
\def\vomega{{\bf \omega}}  %\def\vomega{{\vec \omega}}
\def\vv{{\bf v}}           %\def\vv{{\vec v}}         
\def\mg{\left\langle} % \big <
\def\md{\right\rangle} %\big >

\def\ltsima{$\; \buildrel < \over \sim \;$}
\def\simlt{\lower.5ex\hbox{\ltsima}}
\def\gtsima{$\; \buildrel > \over \sim \;$}
\def\simgt{\lower.5ex\hbox{\gtsima}}

\def\mG{{\cal G}}

\newcount\eqnum\eqnum=0% register
\def\autnum{\global\advance\eqnum by 1\relax{\rm(\the\eqnum)}}

\pagerange{000--000}    % `letters' articles should use \pagerange{Ln--Ln}
\pubyear{1996}
\volume{000}
\title[The $\Omega$ dependence of the velocity divergence
distribution]{The $\Omega$ dependence of the velocity divergence distribution}
\author[Bernardeau et al.]{F. Bernardeau$^1$, R. van de Weygaert$^2$,
E. Hivon$^{3,4}$ and F. R. Bouchet$^4$ \\
$^1$ Service de Physique Th\'eorique, C.E. de Saclay,
F-91191 Gif-sur-Yvette c\'edex, France\\
$^2$ Kapteyn Astronomical Institute, University of Groningen,
P.O. Box 800, 9700 AV Groningen, the Netherlands\\
$^3$ Theoretical Astrophysics Center, Juliane Maries Vej
30, DK-2100 Copenhagen \O, Denmark.\\
$^4$ Institut d'Astrophysique de Paris, CNRS, 98 bis boulevard Arago, F-75014 Paris, France}

\begin{document}
\setcounter{table}{0}
\setcounter{equation}{0}

\maketitle  %  finish the two spanning material

\begin{abstract}
{Analytical studies based on perturbative theory have shown that the
moments of the Probability Distribution Function (PDF) of the local
smoothed velocity divergence are expected to have a very specific
dependence on the density parameter $\Omega$ in the quasi-linear
regime. This dependence is particularly interesting as it does not 
involve the possible bias between the galaxy spatial distribution and
the underlying mass distribution. This implies a new and promising 
method for determining a bias-independent value of $\Omega$ based 
on a reliable determination of the velocity divergence PDF. 

In this paper we study the $\Omega$ dependence of the velocity
divergence PDF and its first moments in a set of N-body simulations,
using the so-called Voronoi and Delaunay methods. We show that this
dependence is in agreement with the theoretical prediction, even while
the number density of velocity field tracers has been diluted to 
a value comparable to that available in current galaxy catalogues.

In addition, we demonstrate that a sufficiently reliable determination
of these statistical quantities is also possible when the measurement of
the galaxy peculiar velocities is restricted to the one component
along the line-of-sight. Under {\it ideal}, noise-free circumstances 
we can successfully discriminate between low and high $\Omega$. 

%However we have not yet investigated the possibility 
%of building a complete procedure based on these numerical methods
%that takes into account realistic sampling and noise.
}
\end{abstract}


\section{Introduction}
The study of the cosmic velocity field is a very promising and
crucial area for the understanding of large--scale structure
formation. Since the early work of Rubin et al. (1976) and Burstein et
 al. (1987) a lot of effort has been invested in the measurement of the
large--scale velocity flows (see Dekel 1994 and Strauss \& Willick
1995 for recent reviews of the subject). Particularly important 
for the analysis of these measured velocity fields has been the 
development of the parameter-free POTENT method by Bertschinger 
\& Dekel (1989). Based on the plausible assumption of potential flow 
it enabled the construction and study of the full three-dimensional 
velocity field in a fair fraction of the local Universe out of the 
measurements of galaxy line-of-sight peculiar velocities. This 
opened up and triggered a host of studies addressing various issues 
and aspects of cosmic velocity flows and provided a versatile 
ground for testing the scenario of large--scale structure formation.

The cosmic velocity field is particularly interesting because of its 
close and direct relation to the underlying field of mass 
fluctuations. Indeed, on these large scales the acceleration, and
therefore the velocity, of any object is expected to have an 
exclusively gravitational origin so that it should be independent of its 
nature, whether it concerns a dark matter particle or a bright
galaxy. Moreover, the linear theory of the generic gravitational 
instability scenario predicts that at every location in the Universe 
the local velocity is related to the local acceleration, and hence the
local mass density fluctuation field, through the same universal 
function $f(\Omega)\approx\Omega^{0.6}$ (Peebles 1980). As the linear 
theory provides a good description on those large scales the 
use of this straightforward relation implies the possibility of a
simple inversion of the measured velocity field into a field that is 
directly proportional to the field of local mass density 
fluctuations $\delta = \rho/\mg\rho\md -1$. %HE --- def of delta
Such a procedure can be used to infer the value of 
$\Omega$, through a comparison of the resulting field with the field of mass 
density fluctuations in the same region. However, the determination of this 
mass density fluctuation field through the measurement of the 
local galaxy density fluctuation field, $\delta_g$, may 
be contrived. The galaxy distribution may be representing a biased
view of the underlying mass density fluctuation field. A common and 
rather simplistic assumption is that $\delta_g$ and $\delta$ are
related via a linear bias factor $b$, 
\begin{equation}
	\delta_g(\vr)=b\ \delta(\vr).
	\label{eq:def_bias}
\end{equation}
However, although several physical mechanisms have been invoked to 
explain such a {\it linear bias model} (see e.g. Dekel \& Rees 
1987), by lack of a complete and self-consistent theory of galaxy 
formation it should as yet only be considered as a numerical 
factor roughly describing the contrast of galaxy density fluctuations 
with respect to the mass density fluctuations. 

The comparison between the observed local galaxy density fluctuation
field and the local cosmic velocity field, invoking
equation~(\ref{eq:def_bias}), will therefore provide an estimate of
the ratio,
\begin{equation}
	\beta=\frac{f(\Omega)}{b}\approx \frac{\Omega^{0.6}}{b}. 
\end{equation}
Various studies, most notably the ones based on a comparison of the 
galaxy density field inferred from the IRAS redshift survey of Strauss
et al. (1990) and the local velocity field reconstructed by the POTENT
algorithm (Bertschinger et al. 1990; Dekel, Bertschinger \& Faber 1990; 
see Dekel 1994), have yielded estimates of $\beta$ in the range 
$\beta \approx 0.5-1.2$ (see Dekel 1994, Strauss \& Willick 1995,
for compilations of results).

%%%%%%%%%=================================================================
\begin{figure}
\vskip 7.5 cm
\special{hscale=50 vscale=50 hoffset=0 voffset=-80 psfile=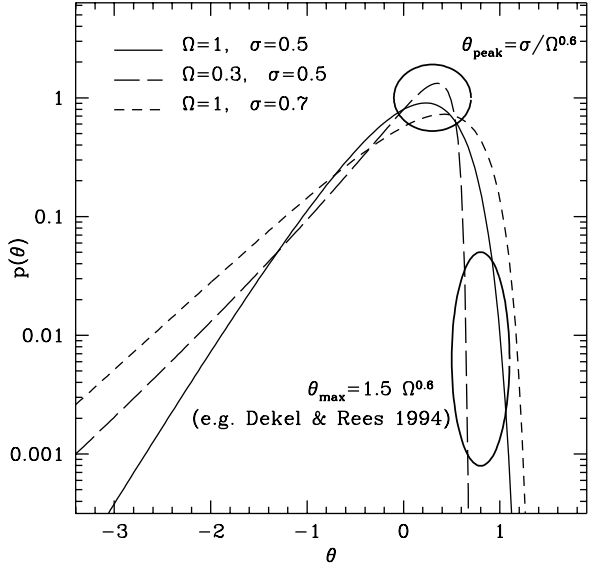}
\caption{ 
The PDF of the velocity divergence as given by Eq. (\ref{eq:PDF_theor_exact})
showing its dependence with $\Omega$ and $\sigma$.}
\label{fig:PDF_theor}
\end{figure}
%%%%%%%%%=================================================================

It is then of crucial interest to find ways to disentangle the
contribution of $\Omega$ and $b$ to $\beta$. A variety of methods using {\it 
intrinsic} properties of the large-scale velocity field have 
been proposed to achieve this. One such attempt is based on the 
reconstruction of the initial density field from the observed
distribution of matter through the use of the Zel'dovich
approximation. The further assumption of Gaussianity of the
initial density probability distribution leads to a constraint on
$\Omega$ (Nusser \& Dekel 1993). Although promising, the quantitative 
results of this approach may be questionable, as the Zel'dovich
approximation provides non-exact results for the induced
non-Gaussian properties of the density and velocity field (Bouchet et
al 1992, Bernardeau 1994a). Another interesting attempt is the one 
by Dekel \& Rees (1994), who exploit the simple observation that
voids have a maximal ``emptiness'' of $\delta=-1$. On the basis 
of the corresponding analysis of the measured velocity field in and 
around a void region these authors inferred a lower limit on $\Omega$ 
of about 0.3. 

In this paper we focus on a method to determine the value of
$\Omega$ that finds its origin in a statistical analysis of the 
velocity field. The foundation for this method is formed by analytical
work within the context of the perturbation theory for the evolution 
of density and velocity fluctuations. In particular, it focuses on
the statistical properties of the divergence of the
locally smoothed velocity field $\theta$, which is defined as
\begin{equation}
	\theta \equiv \frac{\nabla \cdot {\bmit v}}{H},
	\label{eq:def_theta}
\end{equation}
where $\bmit v$ is the time derivative of the comoving coordinate
$\bmit x$, and $\nabla = {\partial}/{\partial \bmit x}$. The method,  
proposed by Bernardeau (1994a) and Bernardeau et al. (1995), 
exploits the relations between the lower order moments of the 
probability distribution function (PDF) of $\theta$, and the 
explicit dependence on $\Omega$ of these relations. The specific 
form of these relations were derived under the assumption of 
Gaussian initial conditions. 

The viability of the method was demonstrated by Bernardeau et
al. (1995), who used the strong $\Omega$-dependence of the skewness
factor of $\theta$, or rather the third normalized moment $T_3$, 
\begin{equation}
      T_3 \equiv \mg\theta^3\md/\mg\theta^2\md^2 \propto
\Omega^{-0.6} 
\end{equation}
to estimate successfully the density parameter in N-body simulations 
of structure formation. A tentative application of this method to the 
observed velocity field as processed by the POTENT method yielded results
consistent with $\Omega=1$. Moreover, subsequent work by Bernardeau \& van de
Weygaert (1996) showed that the theoretical predictions concerning the
complete overall shape of the PDF of $\theta$ were valid. They
demonstrated this by comparison of the analytically predicted PDF with
the PDF determined from a large CDM N-body simulation with
$\Omega=1$. In order to deal with the major complication of obtaining  
clean and straightforward numerical estimates of the statistical 
properties of the velocity divergence field from the discretely 
sampled velocity field, they developed two new techniques. These 
techniques exploit the minimum triangulation properties of Voronoi and
Delaunay tessellations (Voronoi 1908, Delaunay 1934, see Van de 
Weygaert 1991, 1994 for references and applications). 

A Voronoi tessellation of a set of particles is a space filling
network of polyhedral cells, with each cell being defined by 
one of the particles as its nucleus and delimiting the part of space 
closer to this nucleus than to any other of the particles. Closely 
related to the Voronoi tessellation is the Delaunay tessellation, 
a space filling lattice of tetrahedra (in three dimensions). Each of 
the tetrahedra in the Delaunay tessellation has four particles of 
the set as its vertices, such that the corresponding circumscribing 
sphere does not have any other particle inside. Through the duality  
relation of the Voronoi and the Delaunay tessellation it is possible 
to obtain one from the other. 

In the method based on the Voronoi tessellation -- the ``Voronoi method''
-- the velocity field is defined to be uniform within each Voronoi 
polyhedron, with the velocity at every location within each cell being
equal to that of its nucleus. The obvious implication of such an
interpolation scheme is that the only regions of space where the
velocity divergence, as well as the shear and vorticity, acquires a 
non-zero value is in the polygonal walls that separate the cells.  
While the Voronoi method can be regarded as a zeroth order
interpolation scheme, yielding a discontinuous velocity field, the 
``Delaunay method'' can be seen as the corresponding first order 
scheme. Basically it constructs the velocity field within each 
Delaunay tetrahedron through linear interpolation between the
velocities of the four defining particles. Evidently, the velocity 
field constructed by the Delaunay method is a field of uniform 
velocity field gradients within each Delaunay cell. 
In the following we consider the statistical properties of the local
velocity divergence when it is filtered by a top-hat window function
following a {\it volume weighting} prescription.
In practice the local filtered divergence is computed on $50^3$
grid points in all cases. Depending on the method used to define the
velocity field, the filtered divergence is given by
a sum of intersections of a sphere with either planar polygonal walls or
tetrahedra (each of them being multiplied by the local divergence)
divided by the volume of the sphere. In such a volume weighting
scheme the filtered quantities do not depend 
on the local number density of tracers. But of course the more
numerous they are, the more accurately determined are the local divergences.
For an extensive description and preliminary tests of the two techniques 
we refer to Bernardeau \& van de Weygaert (1996). 

With the intention of demonstrating the potential and practical 
applicability of our method, this paper presents the results of 
a systematic study of the $\Omega$ dependence of the moments and 
the PDF of the velocity divergence $\theta$ in N-body simulations 
of structure formation, using the numerical schemes of the Voronoi 
and the Delaunay method. To this end, we will first recall the
relevant theoretical results on the statistical properties of 
$\theta$ in Section \ref{sec:PT}, thereby underlining the main
features of importance to our study. This theoretical groundwork 
is followed by Section \ref{sec:num}, containing the presentation of 
the numerical results of the statistical analysis of PM N-body
simulations of structure formation in $\Omega=1$ and $\Omega<1$ 
universes. While the first subsection, \S~\ref{sec:large_sample},
concerns the statistical quantities that were determined in a very large 
sample and thus with maximum attainable accuracy, the second
subsection \S~\ref{sec:dilut} is devoted to the issue to what extent
these results get affected when the sample is strongly diluted. The 
latter is particularly important as we are interested in the
reliability of our method for samples with a number density comparable
to that of available galaxy catalogues. Also of immediate 
relevance for a practical application is the question of how far 
%HE --- in how far 
the results of the statistical analysis get influenced when the 
velocity of the particles in the sample are known along only one 
direction. This issue, of crucial importance within the context 
of the statistical analysis of observational catalogues, is treated 
in a third subsection, in \S~\ref{sec:1D}. Finally, following the 
successful application of our method under the circumstances 
described above, we conclude with a summary and a discussion of 
possible complications and prospects for our statistical method 
to infer a bias-independent value of $\Omega$. 

\section{Perturbation Theory of Structure Formation and the 
velocity field Probability Distribution Function}
\label{sec:PT}
%Another approach has been proposed by Bernardeau et al. (1995) using
%the low order moment of the distribution of the velocity
%divergence. 
Perturbation Theory (PT) is extremely useful for the study and 
analytical description of the mildly non linear evolution of density
and velocity fields. In particular within the context of structure
forming out of Gaussian initial conditions perturbation theory has 
been extensively developed in a large body of work (see e.g. 
Bernardeau 1994a,b). In the case of these Gaussian initial conditions 
the complete set of moments of the smoothed velocity and density fields 
can be computed analytically, in particular if these fields are 
top-hat filtered. The corresponding PDF can be computed through 
re-summation of the series of moments. 

One of the most straightforward and useful results in the context of 
perturbation theory is the relation between the third moment $\mg\theta^3\md$ 
and the second moment $\mg\theta^2\md$ of the probability distribution
function of $\theta$, 
\begin{equation}
	\mg\theta^3\md=T_3\,\mg\theta^2\md^2 = T_3\,\sigma_{\theta}^4.
\end{equation}
The coefficient $T_3$ depends on the cosmological parameter $\Omega$, 
on the shape of the power spectrum, on the geometry of the window
function that has been used to filter the velocity field and even 
on the value of the cosmological constant $\Lambda$, although the 
latter is an almost negligible weak dependence. In fact, the
dependence of $T_3$ on $\Omega$ is substantially stronger than the
one of the 
equivalent coefficient for density field. For
instance, for a top-hat window function and a power law initial power 
spectrum of index $n$, i.e. 
\begin{equation}
%   P(k) \equiv \mg\delta({\bmit k})^2\md = \mg\theta({\bmit k})^2\md 
%\propto k^{n}),
   P(k) \equiv \mg\delta({\bmit k})^2\md 
\propto \mg\theta({\bmit k})^2\md 
\propto k^{n},
\end{equation}
one obtains the following expression for $T_3$, 
\begin{equation}
	T_3={-1\over \Omega^{0.6}}\left[{26\over7}-(3+n)\right].
\end{equation}
As $T_3$ can be directly determined from observations, one can use its
strong dependence on $\Omega$ to obtain an estimate of $\Omega$, as 
has been done by Bernardeau et al. (1995). 

More generally, Perturbation Theory enables one to infer the whole set of
the cumulants $\mg\theta^p\md$ to their leading order. All of
them are related to the second moment via the relation 
\begin{equation}
	\mg\theta^p\md=T_p\,\mg\theta^2\md^{p-1}, 
	\label{eq:def_Tp}
\end{equation}
and as in the case of $T_3$ all the coefficients $T_p$ possess a
strong dependence on the value of $\Omega$ (Bernardeau 1994b). To a good
approximation, this dependence on cosmological parameters can be 
written as 
\begin{equation}
	T_p(\Omega,\Lambda)\approx{1\over\Omega^{(p-2)\,0.6}}T_p(\Omega=1,\Lambda=0).
\end{equation}
This property, given here for the moments, naturally
extends itself to the shape of the complete velocity divergence 
PDF $p(\theta)$. This can be directly appreciated from the work 
by Bernardeau (1994b), who showed that the PDF can be calculated 
from its moments $T_p$ and the value of $\sigma_{\theta}^2$
through a Laplace transform of its generating function $\varphi\v$
\begin{equation}
	p(\Omega,\theta)\d\theta=\int_{-\ii\infty}^{+\ii\infty}
	{\d y\over 
	2\pi\ii\sigma\v^2}\exp\left[-{ \varphi\v(\Omega,y)\over
	 \sigma\v^2}
	+{ y\theta\over \displaystyle \sigma\v^2}\right] 
	\d\theta\,.
	\label{eq:PDF_theor_exact}
\end{equation}
The moment generating function $\varphi\v(\Omega,y)$, given by 
%HE --- can where
\begin{equation}
	\varphi\v(\Omega,y)=\sum_{p=2}^{\infty}\ -T_p(\Omega) 
\frac{(-y)^p}{p!},
	\label{eq:gener_f}
\end{equation}
can be related to the spherical collapse dynamics in the cosmology
under consideration (see Appendix A). Although this calculation is 
almost intractable in the general case, it has been shown that at 
least in one particular case one can evaluate this expression 
analytically. In the specific case of a power law density perturbation
spectrum with an index $n=-1$ in combination with a top-hat smoothing,
it is possible to invoke some approximations that enable the 
derivation of a simple analytic fit for the PDF 
$p(\theta)$\footnote{This fit is obtained
through approximations which tend to lower the values of the
moments (appendix A).
The PDF (\ref{eq:PDF_theor}) presented here is actually more
accurate if $n\approx -1.3$.} (see Appendix A). This fit is given by
\begin{eqnarray}
	p(\theta){\rm d}\theta &=& {([2\kappa-1]/\kappa^{1/2}+
	[\lambda-1]/\lambda^{1/2})^{-3/2} \over \kappa^{3/4} (2\pi)^{1/2}
	\sigma_{\theta}}\, \nonumber \\
	& \times &
	\exp\left[-{ \theta^2\over 2\lambda 
	\sigma_{\theta}^2}\right]\,{\rm d}\theta, 
	\label{eq:PDF_theor}
\end{eqnarray}
with 
$\kappa=1+\theta^2/\left(9\lambda \Omega^{1.2}\right),$ 
and $\lambda=1-2\theta/\left(3\Omega^{0.6}\right).$

%%%%%%%%%=================================================================
\begin{figure*}
\vskip 16.7 cm
\special{hscale=90 vscale=90 voffset=-150 hoffset=-10 psfile=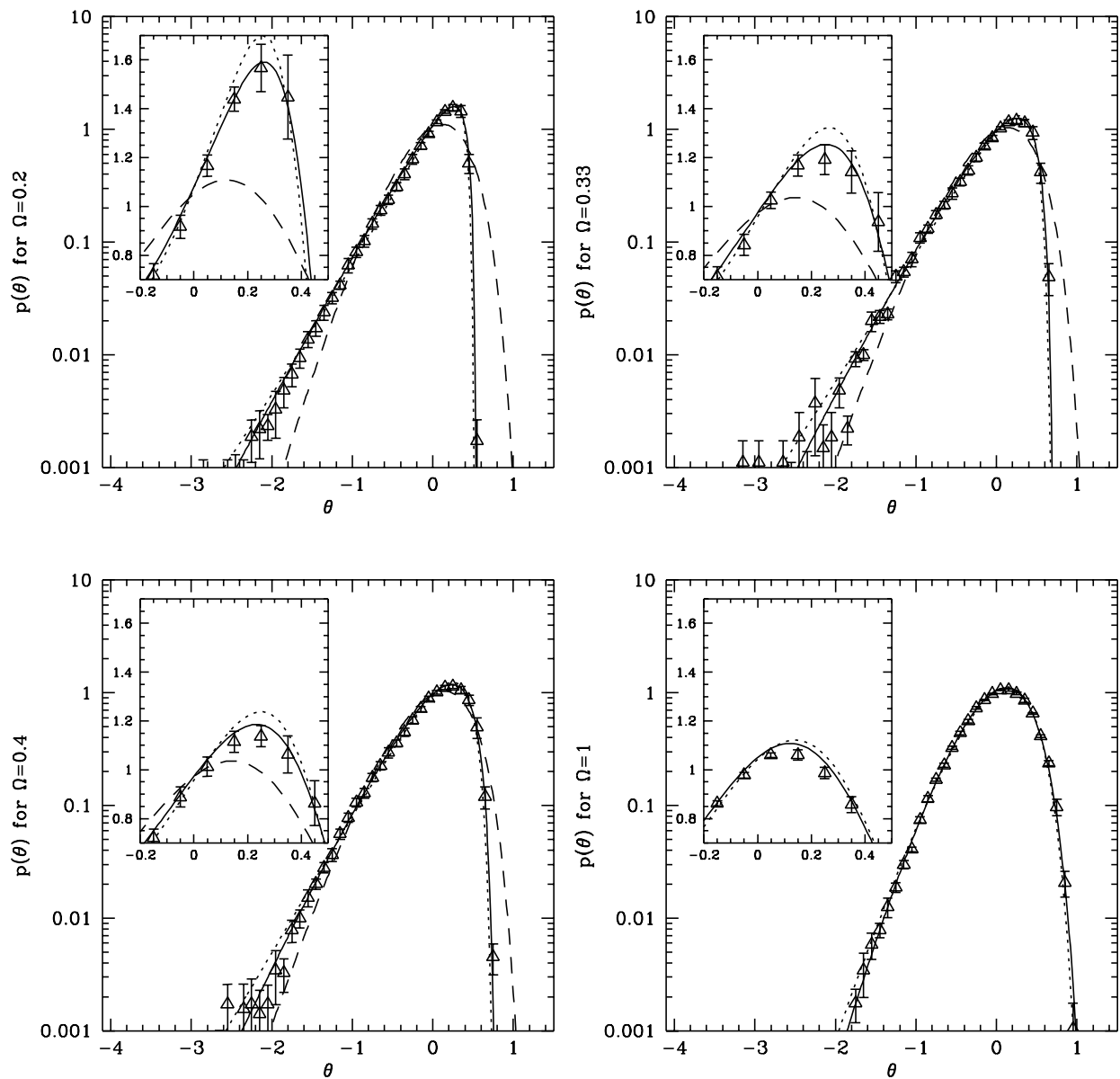}
\caption{The PDF of the velocity divergence for various values
of $\Omega$. The dotted lines correspond to the approximate analytic
fit (Eq.\ref{eq:PDF_theor}) and the solid lines to the theoretical predictions
using (A.15) and (A.14) with $n=-0.7$ obtained for the measured values
of $\sigma$ and $\Omega$. 
The dashed lines are the predictions for $\Omega=1$ and the
same variance. The numerical estimations have been obtained using the
Delaunay method.}
\label{fig:PDF_large_sample}
\end{figure*}
%%%%%%%%%=================================================================
%%%%%%%%%=================================================================
\begin{figure*}
\vskip 7.8 cm
\special{hscale=90 vscale=90 voffset=-390 hoffset=-10 psfile=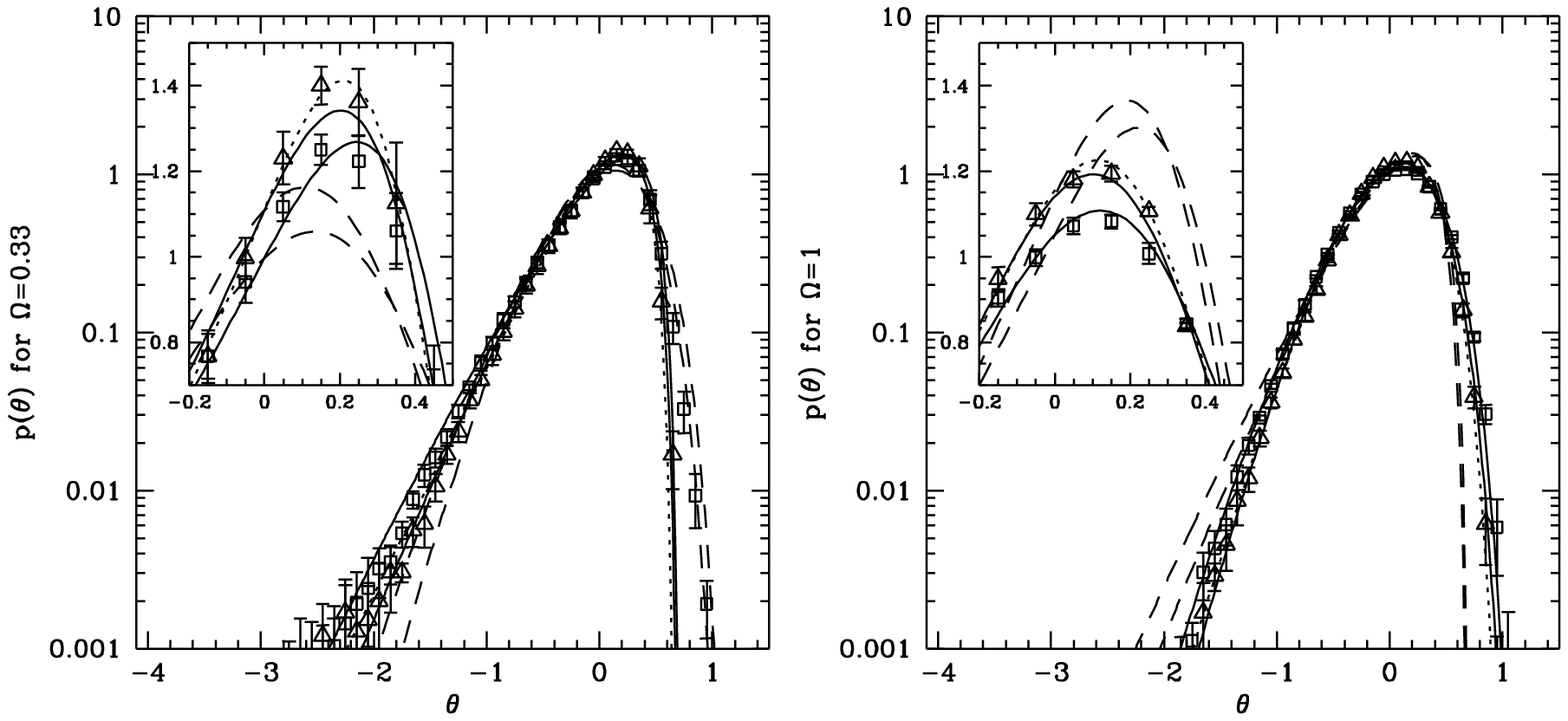}
\caption{Effects of dilution for $\Omega=0.33$ (left panels)
 and $\Omega=1$ (right panels). The PDF-s have been obtained
with only 10,000 tracers in total which gives an average of
about 10 particles per cell. 
The squares show the results of the Voronoi
method and the triangles the results of the Delaunay method.
The solid lines correspond to the theoretical predictions
for the variance obtained with each method and the ``right''
assumption for $\Omega$ and the dashed line with the ``wrong''
assumption for each case and method.}
\label{fig:PDF_small_sample}
\end{figure*}
%%%%%%%%%=================================================================

The behaviour of this function $p(\theta)$ has been illustrated in 
figure \ref{fig:PDF_theor} for various values of $\Omega$ and
$\sigma_{\theta}$. Qualitatively, one can see that the dependence of 
the shape on $\Omega$ reveals itself in two ways: (1) the location of 
its cut-off at high positive values of $\theta$ and (2) the location
of its peak. 

As for (1), the maximum value that $\theta$ can obtain is known exactly 
and is not dependent on the approximations that have been invoked to 
derive expression (\ref{eq:PDF_theor})
\begin{equation}
	\theta_{\rm max}=1.5\ \Omega^{0.6}.
\end{equation}
The value of $\theta_{\rm max}$ determines the location of the cut-off,
and therefore the maximum expansion rate in voids. The value of 1.5 is
the difference in value of the Hubble parameter in an empty,
$\Omega=0$, Universe and that of an Einstein-de Sitter Universe, 
$\Omega=1$. Evidently, this is reflecting the fact that the interior of 
the deepest voids locally mimic the behaviour of an $\Omega=0$
Universe. 
%HE Recall that also the suggestion by Dekel \& Rees (1994) of 
Recall that the suggestion by Dekel \& Rees (1994) of 
using the maximum emptiness of voids to constrain $\Omega$ is also %HE
based on a similar feature. 

Also quite sensitive to the value of $\Omega$ are the position of the
peak of the distribution function $p(\theta)$, i.e. the most likely
value of $\theta$, and the overall shape of $p(\theta)$. Using the Edgeworth
expansion (Juszkiewicz et al. 1995, Bernardeau \& Kofman 1995)
one can show that the value of $\theta$ for which the distribution
reaches its maximum is given by
\begin{equation}
	\theta_{\rm peak}\approx
	-{T_3\over 2}\ \sigma_{\theta}={1\over \Omega^{0.6}} \sigma_{\theta}.
\end{equation}
In fact, a procedure exploiting this dependence of shape and peak
 location of $p(\theta)$ will probably yield a more robust measure of 
$\Omega$ than the maximum value of $\theta_{\rm max}$ as it will be 
less bothered by the noise in the tails.

\section{The $\Omega$ dependence in numerical simulations}
\label{sec:num}
By means of numerical simulations we have investigated the discussed 
dependence of the PDF of $\theta$. These N-body simulations use a 
Particle-Mesh (PM) code (Moutarde et al. 1991) with a $256^3$ grid to 
follow the evolution of a system of $256^3$ particles. For our 
project we used two simulations, one with $\Omega$ having a value of 
$\Omega=1$ and the second one of $\Omega < 1$. By analyzing the latter
at different time-steps we explore situations for different 
values of $\Omega$. The particle distribution in the two simulations 
corresponds to a density and velocity fluctuation field with a
$P(k)\propto k^{-1}$ spectrum. 

As can be seen in Table~\ref{tab:cumulant}, the variances
$\sigma_{\theta}$ do not differ significantly for the different values
of $\Omega$ for a given filtering radius. The fact that the values of
the variance are comparable simplifies a comparison of the PDF
substantially, which makes the interpretation 
in terms of the intrinsic $\Omega$ dependence more straightforward. 

\subsection{Measurements with a large number of tracers}
\label{sec:large_sample}
The first step of our analysis concerns an exploration of the velocity
field using a large number of tracers. For this
study the number of selected particles in each simulation is about
70,000, which for a cell radius of about 6\% of the box size leads to
a mean number of 67 particles per cell.  The selection procedure used
here is deliberately biased towards low-density regions by inducing it
to retain a uniform density of particles all over the simulation
box. Except for its goal of achieving a better velocity
field coverage of low-density regions such a selection bias is not
expected to influence the velocity field analysis.  

The methods that we use to analyze the simulations are exactly the 
same ones as described by Bernardeau \& van de Weygaert
(1996). In fact, at this stage we only used the Delaunay method 
to calculate numerically the shape of the PDF of the velocity 
divergence. The results for the case of a large number of 
velocity field tracers are shown in Fig.~\ref{fig:PDF_large_sample}.   
The results are in good agreement with the theoretical predictions for
the values of $T_3$ and $T_4$ (see Table~\ref{tab:cumulant}), as well
as with the theoretical shape of the PDF (Fig.~\ref{fig:PDF_theor}).

It is in particular worth noting that the specific features expected
from equation~(\ref{eq:PDF_theor}) are indeed confirmed by the
numerical results. Notably, the locations of the cut-off, which are
very sensitive to rare event discrepancies, are 
well reproduced (solid and dotted lines). 
Moreover, as can be observed from the insets,  
also the position and shape of the peak have been reproduced 
very well (solid lines), 
providing a strong discriminatory tool between different values of $\Omega$.

Within the context of these observations, we should issue a few 
side remarks. Although the shape (\ref{eq:PDF_theor}) 
is very attractive because
it is a close analytic form, one should have in mind that it is only 
approximate. Indeed it is derived from an approximate expression
for the cumulant generating function. The differences do not reveal
for the overall shape (the logarithmic plots) but are significant
for the shape of the peaks. For calculating the theoretical
predictions  we are then forced to use a more accurate description
of the cumulants. To achieve this we use the relations (A.13, A.14)
with $n=-0.7$ (the expression \ref{eq:PDF_theor}
corresponds to $n=-1$) in the integral (A.15) which is then 
computed numerically.
It is still an approximate expression, but
it yields the correct value for $T_3$ and a very good
approximation for the higher order cumulants. We should emphasize
that this slight
modification is only instrumental in obtaining the correct shape of the PDF
around its maximum, which is indeed almost entirely
determined by the values of the low order moments. This may be 
understood for example from the properties of the Edgeworth expansion, for 
which we refer to Juszkiewicz et al. 1995 and Bernardeau \& Kofman 1995.

\subsection{The effects of dilution}
\label{sec:dilut}
In order to check the robustness of the results when only
a limited number of tracers for the velocity field is available, 
we performed numerical experiments where only 10,000 particles are 
used to trace the velocity field. The selection of the sample 
points in this diluted sample is completely random and does not
invoke the specific biased selection procedure that was used in 
the case described in the former subsection. For this case of diluted 
samples, we used both the Delaunay and the Voronoi methods for 
analysis. 

Figure~\ref{fig:PDF_small_sample} shows the PDFs obtained with both
methods, for an $\Omega=1$ simulation and for an $\Omega=0.33$
simulation. Both the Voronoi and the Delaunay methods appear to 
yield numerical results that are in reasonably good agreement with the
predicted PDF. Particularly encouraging is 
the result born out by the insets, namely the fact that the shape of 
the peak can still be used as a strong discriminatory tool between different
values of $\Omega$.

Investigating Fig.~\ref{fig:PDF_small_sample} in somewhat more detail,
we can observe that the results obtained by the two methods are affected
somewhat differently by the dilution procedure. The PDFs obtained with the 
Voronoi method generally possess a less sharply defined tail at the
side of the high $\theta$ values. This effect appears to be stronger for the 
low $\Omega$ case. This behaviour may originate in the fact 
that the divergence is localized to a limited part of space, non-zero 
values being confined to the walls of the tessellation. In such a 
situation Poisson like errors in the measurements are expected to
become particularly prominent in the heavily diluted areas of the void
regions. An additional difference is a slight underestimation of the 
values of the $T_p$ coefficients by the Voronoi method (see 
Table~\ref{tab:cumulant}). Unfortunately, we do not see a possibility 
to correct for such effects. On the other hand, the Delaunay method
seems to be more robust against such effects. However, it tends to 
underestimate the value of the variance and the higher order moments. 
The latter is probably a consequence of the fact that the effective
filtering radius tends to be larger.

\subsection{The effects of reducing the information to one velocity component}
\label{sec:1D}
A major complication in the analysis of velocity fields under practical 
circumstances is the fact that only the velocity component along the 
line of sight can be measured. This therefore forms the second issue
that we address in this paper. 

As yet we restrict ourselves to an artificial situation with ideal 
measurements, not yet to investigate the determination of the statistical 
quantities associated with the velocity field under realistic 
circumstances. To simplify furthermore our investigations we assume
that we can use
 the approximation of an infinitely remote observer so that the radial
velocity can be identified with one velocity component, namely
the $x$-direction in the following.
More specifically, we address the effects of the
reduction of information concerning the velocity field traced by 
a diluted sample of points. In principle, the fact that the velocity 
field is only known along one direction should not pose any problem. 
In the usual structure formation scenarios based on gravitational 
instability the large scale velocity field is expected to be 
non-rotational, implying it to be a potential flow and therefore 
the gradient of a potential that can be inferred from the 
measurement of only one component of the velocity (Bertschinger et
al. 1990, Dekel et al. 1990).

%%%%%%%%%=================================================================
\begin{figure*}
\vskip  7.8 cm
\special{hscale=90 vscale=90 voffset=-390 hoffset=-10 psfile=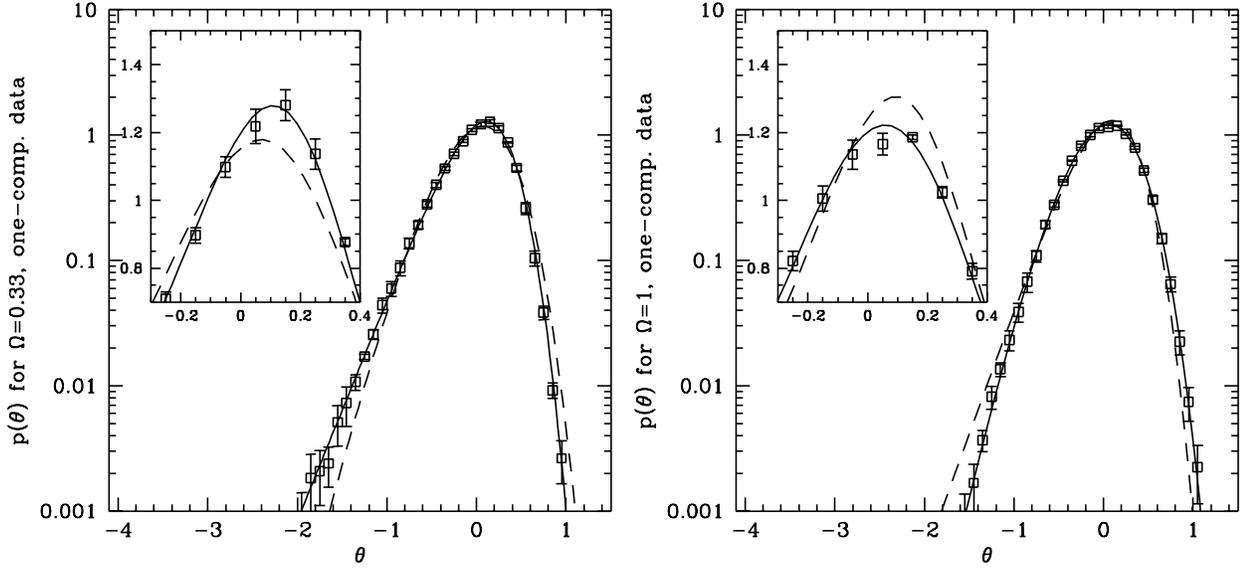}
\caption{Effects of reduction of the information on the velocity 
field to only one component for $\Omega=0.33$
(left panel) and $\Omega=1$ (right panel). We used the estimator
(\ref{eq:theta_w_estim}) with $\epsilon=0.1$ to determine
numerically the local divergences.
The solid lines correspond to the model (\ref{eq:PDF_1D_estim}) 
obtained from (\ref{eq:theta_estim_model})
with the parameters $\mu$ and $\sigma_e$ given  in Table~\ref{tab:par1D}.
The dashed lines are the resulting shapes of the PDF-s when one
uses the same parameters but the ``wrong'' value for $\Omega$
($1$ instead of $0.33$ in the left panel, $0.33$ instead of $1$ in the
right one).}
\label{fig:PDF_1D}
\end{figure*}
%%%%%%%%%=================================================================

%%%%%%%%%=================================================================
\begin{figure*}
\vskip  11.5 cm
\special{hscale=90 vscale=90 voffset=-300 hoffset=-30 psfile=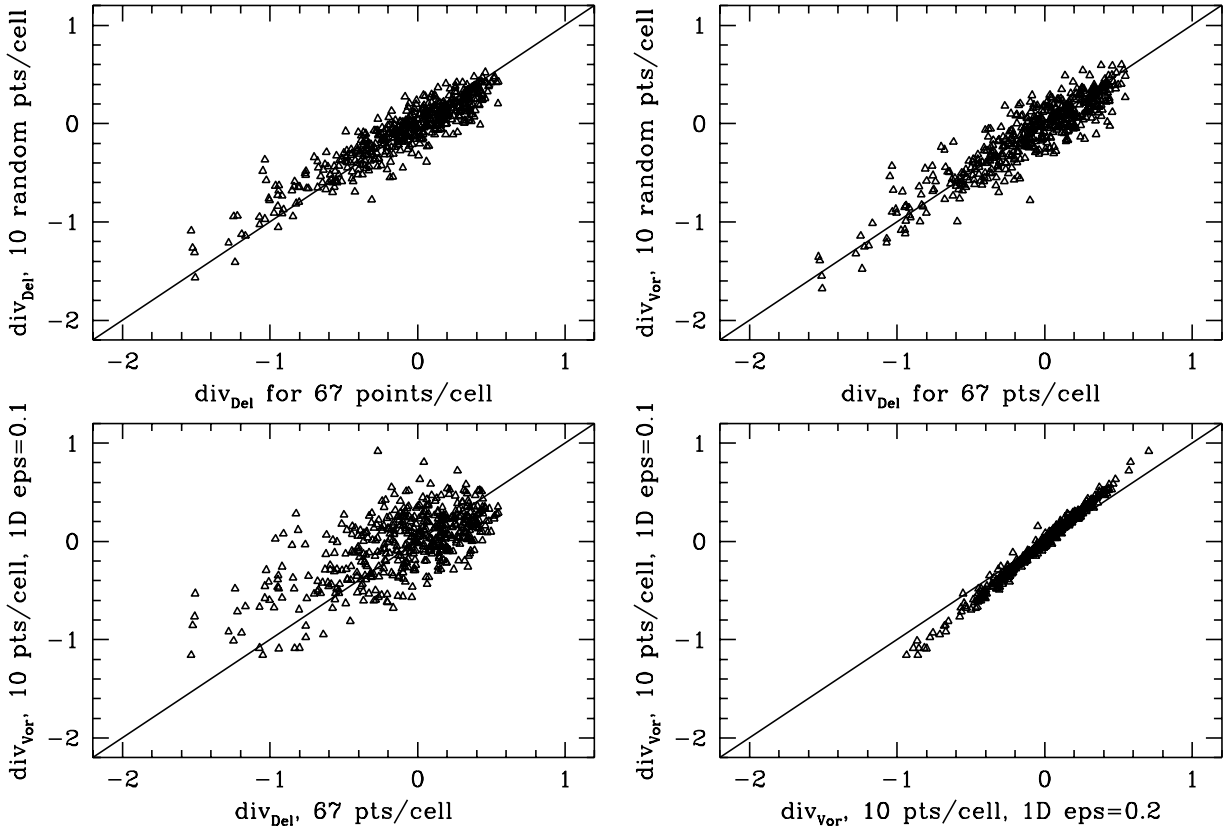}
\caption{Scatter plots that show the differences of the various
estimators of the local divergence measured at 500 different
random locations. }
\label{fig:scatt_div}
\end{figure*}
%%%%%%%%%=================================================================

In the Voronoi method non-zero values of the divergence $\theta$ are 
restricted to the walls of the Voronoi tessellation, where the local 
divergence is given by 
\begin{equation}
	\theta_{\rm wall}=\vn\,.\,\Delta \vv,
	\label{eq:theta_w_exact}
\end{equation}
with $\vn$ being the normed vector orthogonal to the wall and
$\Delta \vv$ the difference of the velocities on the opposite sides 
of the Voronoi wall. The expression for the local vorticity is given by
\begin{equation}
	\vomega_{\rm wall}=\vn\,\times\,\Delta\vv.
	\label{eq:vortic_w_exact}
\end{equation}
Assuming potential flow, and hence a zero value of $\vomega$, this 
implies relations between the various components of $\Delta \vv$ and 
the following expressions for $\theta_{\rm wall}$:
\begin{equation}
	\theta_{\rm wall}={\Delta v_x\over n_x}={\Delta v_y\over n_y}=
	{\Delta v_z\over n_z}.
\end{equation}
In practice, this introduces the numerically unstable operation of
dividing by one component $\vn$ as it can be arbitrarily close to
zero. It may therefore be more reasonable to try to estimate the 
value of $\theta_{\rm wall}$ using the stable, but ad-hoc,
prescription of
\begin{equation}
	\theta_{\rm wall}^{\rm estim.}=
	\Delta v_x { n_x\over n_x^2+\epsilon},
	\label{eq:theta_w_estim}
\end{equation}
where $\epsilon$ is a small parameter of the order of $\epsilon
\approx 0.1$. 

Note that such a prescription is not self-consistent in reproducing
the full 3D velocity field.\footnote{This can 
be readily appreciated from the fact that the normal $\vn$ to a wall 
in the Voronoi tessellation is proportional to the vector $\Delta\vr$ 
between the two points on each side of the wall. 
Thus, if it were possible to build a consistent velocity filed from the
constraints (17) it would imply $\Delta\vv\propto
\Delta\vr$ yielding not only a vanishing vorticity
but also a vanishing shear.}Indeed, depending on the way one goes from 
cell A to cell B -- and there are infinitely many ways to go from A to
B, even while they are direct neighbours -- one will not necessarily 
find the same value for $\theta_{\rm wall}^{\rm estim.}$ using the 
equivalent prescriptions for $\Delta v_y$ or $\Delta v_z$. 
The method that we adopt here will 
therefore certainly not be the most accurate method. However, we may 
expect it to be a reasonable approximation, and will therefore use 
it here to illustrate the properties in which we are interested. 

Figure~\ref{fig:PDF_1D} displays the results obtained with
(\ref{eq:theta_w_estim}) for $\epsilon=0.1$. Evidently, the results 
get affected to quite some extent by the transition from 
(\ref{eq:theta_w_exact}) to (\ref{eq:theta_w_estim}). This leads to 
the key question whether it is still feasible to reliably 
recover the statistical information on $\theta$ or not. As can be 
observed from the scatter plots in Fig.~\ref{fig:scatt_div}, the 
change from a situation in which one has knowledge of the full 
velocity to one where this has been limited to only one 
component thereof introduces a large scatter. However, we found that 
it is possible to define a meaningful representation of 
the scatter plots in terms of the following empirical description: 
\begin{equation}
	\theta_{\rm estim.}=\mu\ (\theta+e)\,.
	\label{eq:theta_estim_model}
\end{equation}
Within this expression the coefficient $\mu$ is a constant with some
fixed value. The scatter is represented by the quantity $e$, a
Gaussian random variable whose value is {\em independent} of $\theta$ 
and which has a vanishing mean. Using
the measured values of the variance and the skewness of the
distribution it is possible to estimate the value of $\mu$ and the
value of the rms fluctuation $\sigma_e$ of $e$. From 
(\ref{eq:theta_estim_model}) one can readily infer that 
\begin{eqnarray}
\sigma_{\rm estim.}&=&\mu\ \sqrt{\sigma\v^2+\sigma_e^2} \nonumber \\
T_{3\rm\ estim.}&=&{\sigma\v^4\over \mu\,(\sigma\v^2+\sigma_e^2)^2}\ T_3
\end{eqnarray}
where $\sigma\v$ is the exact rms fluctuations of $\theta$ and $T_3$ 
its third cumulant, while $\sigma_{\rm estim.}$ and $T_{3\rm\ estim.}$
are the corresponding estimated values. By solving this set of
equations one can find the values of $\mu$ and $\sigma_e$. Their
values for various $\Omega$ and $\epsilon$ are listed in 
Table~\ref{tab:par1D}. 

Moreover, significant within the context of the ultimate goal of 
developing an unbiased estimator of $\Omega$, is that these 
parameters were found to be almost independent of the value of 
$\Omega$. More specifically, it turns out that the value of $\mu$ 
only depends on the adopted value of $\epsilon$, whereas $\sigma_e$ 
is independent of $\Omega$ and only marginally dependent on
$\epsilon$. This can clearly be appreciated from the bottom right-hand 
panel of Fig.~\ref{fig:scatt_div}, which demonstrates that the two 
estimations of $\theta$, one based on $\epsilon=0.1$ and the other 
on $\epsilon=0.2$, are basically proportional to each other. 
It may therefore be argued that it is quite natural to expect that 
the noise $e$ introduced by this method is somehow intrinsic to 
the distribution.

In Appendix B we describe an extremely simple model based on the
assumption that the relative velocity of two particles is proportional
to their relative position. This allows us to compute analytically the
parameters $\mu$ and $\sigma_e$ entailed by the use of the numerical scheme
(\ref{eq:theta_w_estim}). One can show that they are both independent on
$\Omega$, and only depend on $\epsilon$ and $\sigma\v$, with $\sigma_e
< \sigma\v$. These analytical predictions are listed in
Table~\ref{tab:par1D_anal}. They appear to be fairly close to their
numerical measurements. The discrepancy 
between the analytical and numerical estimations of these parameters is
due to the loss of information associated with the projection of the 
velocity from three to one dimensions. Although this is not fairly 
represented by our model, the discrepancy seems to be quite small.

%It is probably due to the algorithm used which
%misrepresents the divergence by coupling the divergence and shear components.

%%%%%%%%%=================================================================
\begin{table*}
\begin{minipage}{140mm}
\caption{Cumulants from the Perturbation Theory
and as estimated by the various numerical methods. \hfill}
\label{tab:cumulant}
      \halign{\quad\hfil#\hfil\quad&
\quad\hfil$#$\hfil\quad&\quad\hfil$#$\hfil\quad&\quad\hfil$#$\hfil\quad&\quad\hfil$#$\hfil\quad&\quad\hfil$#$\hfil\quad\cr
            \noalign{\hrule}
            \noalign{\medskip}
 \# tracers per cell & {\rm cumulants} & \Omega=0.2 & \Omega=0.33 & \Omega=0.4 & \Omega=1 \cr
            \noalign{\medskip}
            \noalign{\hrule}
            \noalign{\medskip}
 $\infty$ & T_3^{\rm PT} & -4.5 & -3.33 & -2.97 & -1.71 \cr
 Perturbation Theory  & T_4^{\rm PT} & 31.41 & 17.22 & 13.67 & 4.55 \cr
            \noalign{\medskip}
            \noalign{\hrule}
            \noalign{\medskip}
$\approx 65$& \sigma^{\rm Del} & 0.37\pm0.01& 0.41\pm0.01& 0.42\pm0.01&0.38\pm0.005\cr
 Delaunay Method     & T_3^{\rm Del}    & -4.52\pm0.2& -3.3\pm0.2 & -2.9\pm0.2 &-1.54\pm0.1\cr
          & T_4^{\rm Del}    & 31.5\pm2.5 & 17.3\pm3   &12.8\pm2.5  &3.1\pm0.7\cr
            \noalign{\medskip}
            \noalign{\hrule}
            \noalign{\medskip}
$\approx 10$& \sigma^{\rm Del} & - & 0.36\pm0.01& - & 0.35\pm0.005\cr
Delaunay Method & T_3^{\rm Del}    & - & -3.8\pm0.2 & - &-1.84\pm0.1\cr
          & T_4^{\rm Del}    & - & 24.7\pm3   & - &5.4\pm1\cr
            \noalign{\medskip}
            \noalign{\hrule}
            \noalign{\medskip}
$\approx 10$& \sigma^{\rm Vor} & - & 0.40\pm0.01& - & 0.38\pm0.003\cr
Voronoi Method & T_3^{\rm Vor}    & - & -3.2\pm0.2 & - & -1.51\pm0.04\cr
          & T_4^{\rm Vor}    & - & 19.4\pm5.5 & - & 3.62\pm1.5\cr
            \noalign{\medskip}
            \noalign{\hrule}
            \noalign{\medskip}
$\approx 10$& \sigma^{\rm Vor} & - & 0.35\pm0.006& - & 0.34\pm0.009\cr
one velocity & T_3^{\rm Vor}    & - & -2.40\pm0.3& - & -1.14\pm0.07\cr
component, $\epsilon=0.1$ & T_4^{\rm Vor}    & - & 11.9\pm4 & - & 2.9\pm2.3\cr
            \noalign{\medskip}
            \noalign{\hrule}}
\end{minipage}
\end{table*}
%%%%%%%%%=================================================================

%%%%%%%%%=================================================================
\begin{table}
      \caption{Numerical values of the
parameters $\mu$ (bias) and $\sigma_e$ (noise) introduced by the
calculation (\ref{eq:theta_w_estim}) of the velocity divergence.}
\label{tab:par1D}
\centering
\begin{tabular}{lcc}
\hline
% REVISED VALUES 19/3/97
\ $\Omega=0.33$\ &\ $\epsilon=0.1$ &\ $\epsilon=0.2$\ \\
\hline
\ $\mu$      & 0.73 & 0.56 \\ % FB calculation
\ $\sigma_e$ & 0.25 & 0.27 \\ % FB
\hline
\ $\Omega=1$\ &\ $\epsilon=0.1$ &\ $\epsilon=0.2$\ \\
\hline
\ $\mu$      & 0.75 & 0.57 \\
\ $\sigma_e$ & 0.24 & 0.26 \\
\hline
\end{tabular}
\end{table}
%%%%%%%%%=================================================================

On the basis of the simple model described in 
Eq.~(\ref{eq:theta_estim_model}) it is possible to reconstruct the 
corresponding shape of the distribution of $\theta_{\rm estim}$, 
\begin{equation}
	p_{\rm estim.}(\theta_{\rm estim.})= 
	\int_{-\infty}^{+\infty} 
	p({\theta_{\rm estim.} \over \mu}-e)\ {\exp(-e^2/2/\sigma_e^2)\over
	(2\,\pi)^{1/2}}\,{\d e\over\sigma_e}
	\label{eq:PDF_1D_estim}
\end{equation}
where $p(\theta)$ is given by Eq.~(\ref{eq:PDF_theor}). A comparison 
of the resulting distribution (\ref{eq:PDF_1D_estim}) with the 
measured histograms is shown in Fig.~\ref{fig:PDF_1D}. The agreement 
appears to be quite good, rendering this phenomenological description 
a quite valuable one. 

%%%%%%%%%=================================================================
\begin{table}
      \caption{Analytical estimations of the
parameters $\mu$ and $\sigma_e$ (see Appendix B).}
\label{tab:par1D_anal}
\centering
\begin{tabular}{lcc}
\hline
\ $\Omega=0.33$\ &\ $\epsilon=0.1$ &\ $\epsilon=0.2$\ \\
\hline
\ $\mu(\epsilon)$                 & 0.60 & 0.49 \\
\ $\sigma_e(\epsilon,\sigma\v)\ $ & 0.20 & 0.23 \\
\hline
\ $\Omega=1$\ &\ $\epsilon=0.1$ &\ $\epsilon=0.2$\ \\
\hline
\ $\mu(\epsilon)$                 & 0.60 & 0.49 \\
\ $\sigma_e(\epsilon,\sigma\v)\ $ & 0.19 & 0.22 \\
\hline
\end{tabular}
\end{table}
%%%%%%%%%=================================================================

Evidently, we are still able to clearly distinguish between 
the scenario with a high value of $\Omega$ and the ones with a low 
%HE value, although the distinction is not as clear than in the previous 
    value, although the distinction is not as clear as in the previous 
cases based on ideal sampling circumstances. Looking into some detail,
we come to the conclusion that the signal has been diluted somehow by 
the noise $e$, and that 
there is a competition between the true rms of the divergence 
$\sigma\v$ and the value of $\sigma_e$. In this respect it is also 
important to note that it is crucial for a successful determination 
of $\Omega$ to have good estimates of both $\mu$ and $\sigma_e$. 
Once these parameters have been determined, and as long as 
$\sigma\v > \sigma_e$, we can see no further problem in distinguishing 
between the different cosmological scenarios. Fig.~\ref{fig:PDF_1D} %HE\ref{fig:scatt_div}
gives an idea of the magnitude 
of the discrepancy that one gets when one assumes a wrong value 
of $\Omega$. In the left panel this concerns the case wherein one 
take a value of $\Omega=1$ instead of the actual value of
$\Omega=0.33$, while in the right-hand panel it concerns the reverse 
situation. 

At this point it is worthwhile to stress once more that so far we only 
tested the method for {\em ideal} measurements of the line-of-sight
velocities. Noise in the measured values of these velocities was 
%HE not yet taking into account. Moreover, besides the fact that this 
    not yet taken into account. Moreover, besides the fact that this 
noise has quite a large value, an evaluation of its influence is 
substantially complicated as it not only concerns random measurement 
errors but also contains contributions from a plethora of, partially 
un-understood, systematic effects. In a work in preparation
we will attempt to develop specific techniques for estimating the 
velocity divergence PDF from such noisy line-of-sight measurements.

\section{Discussion and Conclusions}
\label{sec:conclusion}
We have tested and confirmed the validity of the strong $\Omega$
dependence of the Probability Density Function (PDF) of the velocity 
divergence that had been predicted on the basis of analytical
Perturbation Theory calculations. These tests are based on a numerical 
analysis of N-body simulations of structure formation in a Universe 
with a Gaussian initial density and velocity fluctuation field.
On the basis of this verification we may conclude that the analytical 
predictions of Perturbation Theory yield very accurate results for 
a wide range of cosmological models. 

The main practical implication of our work is the basis it offers 
for a potentially very valuable and promising estimator of the 
value of $\Omega$, an estimator independent of a possible bias 
between the distribution of galaxies and the underlying matter 
distribution. The successful tests presented in this work demonstrate 
the validity of the equations of Perturbation Theory that form 
a basis for the estimates of $\Omega$ which are based on the statistical 
properties of the velocity divergence field. Their unbiased nature 
finds its origin in the fact that the relations between the various 
statistical moments do not contain any explicit dependence on a 
bias between the galaxy and the matter distribution. 
Moreover, the estimated values of $\Omega$ are even more 
direct and straightforward to interpret as the relevant statistical 
relations only involve a very weak dependence on the cosmological constant 
$\Lambda$ is expected to be very weak (Bernardeau 1994a, b).

Not only relations between the statistical moments of the $\theta$ 
distribution, also the shape and general functional behaviour provide 
a useful indicator for the value of $\Omega$. When we focus on the 
details of this functional behaviour of the velocity divergence
distribution function -- illustrated in Fig.~\ref{fig:PDF_large_sample} and
\ref{fig:PDF_small_sample}) -- we can draw a few conclusions with
regards the practical feasibility of obtaining reliable estimates 
from the shape of $p(\theta)$. Both location and
shape of the peak of this distribution appear to be robust indicators 
of the value of $\Omega$. On the other hand, the location of the 
maximum of the divergence $\theta$ -- i.e. the cutoff value of
$p(\theta)$ -- appears to be much more sensitive to a poor sampling. 
This may make it harder for it to provide reliable estimates 
of $\Omega$ from currently available observational catalogues 
(see Fig.~\ref{fig:PDF_small_sample}). However, it is all the more 
encouraging that even on the basis of the cutoff value we obtained 
a reasonable agreement between theoretical predictions and numerical 
measurements.

Finally, we addressed one further crucial issue towards an application
of our estimation procedures to real data sets. This concerns the 
problem of not being able to obtain directly the full
three-dimensional velocity field. Instead, the velocity of a galaxy 
can only be measured along the line-of-sight. In a preliminary attempt
to study the consequences of this fact for the feasibility of 
our method, we introduced a partially empirically defined extension 
of our method. Despite of the extreme crudeness and rather ad-hoc 
nature of this algorithm to reconstruct the full velocity field, it 
is quite encouraging that we are able to distinguish between 
the velocity PDF obtained in a flat Universe and that obtained in 
an open Universe. The major obstacle towards a successful application 
of our methods therefore appears to be the one of noisy data sets
and systematic sampling errors.
We have not yet dealt with these 
problems, deferring them to a forthcoming paper. 

An additional and useful application of our numerical work involves 
a test for structure indeed  having emerged through the process of 
gravitational growth of an initially Gaussian
random density and velocity field. Having shown Perturbation Theory 
to be valid, we can exploit its prediction that the PDF of $\theta$ 
is only dependent on a few parameters, in particular $\sigma\v$ and 
$\Omega$. If no values of $\sigma\v$ and $\Omega$ can be found to
produce an acceptable fit to the observed velocity field, this will 
force us to conclude that it is unlikely that the structure 
developed as described within the standard framework of gravitational 
instability and Gaussian initial conditions. In this context it 
is interesting to point out that a negative skewness has been observed
in the currently available datasets (Bernardeau et al. 1995), which is
an indication in favour of standard scenarios.

Summarizing, we may conclude that the combined machinery of the
analytical perturbation theory results and the developed numerical
methods and their application on the intrinsic statistical properties of the
velocity field provides us with a reliable new estimator of the
cosmological density parameter $\Omega$. This estimator is all the 
more useful as it is one of the very few which will yield values 
of $\Omega$ completely independent of galaxy-density field biases 
and almost independent of the value of $\Lambda$. 

\section*{Acknowledgments}
F. Bernardeau would like to thank IAP, where a large part of the work
has been completed, for its warm hospitality. We would like to thank
A. Dekel for encouraging comments and discussions. FB and RvdW are
grateful for the hospitality of the Hebrew University of Jerusalem,
where the last part of this contribution was finished. 

R. van de Weygaert is supported by a fellowship of the Royal
Netherlands Academy of Arts and Sciences. Part of this work was done 
while EH was at the Institut d'Astrophysique de Paris (CNRS),
supported by the Minist\`ere de la Recherche et de la Technologie.
Additional partial support to EH was provided by the Danish National Research
Foundation through the establishment of the Theoretical Astrophysics
Center. The computational means were made available to us thanks to the
scientific council of the Institut du D\'eveloppement et des Resources
en Informatique Scientifique (IDRIS).

%-----------------------------------------------
\appendix

\section[]{Calculation of The Probability Distribution Function of
$\theta$} 
In the case of a top-hat window function it is possible to
evaluate the whole series of cumulants of $\theta$ at their leading
order. This makes it feasible to construct, at least in principle, 
the complete PDF of the local top-hat smoothed velocity
divergence field. However, in order to construct a convenient 
closed form of the PDF we to introduce some approximations. Here we 
will go through some of the analytical results and applied
approximations that were used in order to obtain simplified final expressions.

A key element in the construction of the PDF of $\theta$ is 
the moment generating function $\varphi\v$ and the close relation 
that was discovered to exist (see e.g. Bernardeau 1992) between 
this function and the dynamics of spherical collapse in the 
background Friedmann-Robertson-Walker universe. More specifically, the 
generating function $\varphi\v(\Omega,y)$, defined as 
\begin{equation}
	\varphi\v(\Omega,y)=\sum_{p=2}^{\infty}\ -T_p(\Omega) \frac{(-y)^p}{p!},
	\label{eq:gener_f_APPEND}
\end{equation}
where $T_p$ are the reduced cumulants of $\theta$ introduced in
(\ref{eq:def_Tp}), is determined from a function $\mG\v(\Omega,\tau)$
whose behavior can be deduced from spherical collapse dynamics in the
cosmology under consideration. These two functions
$\varphi\v(\Omega,y)$ and $\mG\v(\Omega,\tau)$ are related through the
system of equations
\begin{eqnarray}
	\varphi\v(\Omega,y) &=& y\mG\v(\Omega,\tau)-
	{1 \over  2}y\tau\, { \d \over  \d \tau}
	\mG\v(\Omega,\tau),\\
	\tau &=& -y\,{ \d \over  \d \tau}\mG\v(\Omega,\tau).
	\label{eq:phi_G_relat}
\end{eqnarray}
For the solution of this system it is therefore necessary to first
determine the explicit relationship of $\mG\v(\Omega,\tau)$ to the
dynamics of spherical collapse. In order to accomplish this, it is
useful to introduce the functions $\mG^{SC}_{\dta}(\tau)$ and
$\mG^{SC}\v[\Omega(a),f(\Omega,\Lambda)\tau]$. The first one of these,
$\mG^{SC}_{\dta}(\tau)$, is defined to be the nonlinear density
contrast of a spherical perturbation of initial over-density $-\tau$.
For exact analytical expressions for $\mG^{SC}_{\dta}$, which exist
only if $\Lambda=0$, we refer to appropriate textbooks. Directly
related to the expression for the density contrast is the function
describing the local departure from the Hubble expansion,
\begin{eqnarray}
\mG^{SC}\v[\Omega(a),f(\Omega,\Lambda)\tau] =
-a{\d \over  \d
a}\mG^{SC}_{\dta}(a,\tau)+ \quad \quad \quad& &  \nonumber \\
f(\Omega,\Lambda) \,
\tau { \d\over  \d \tau}\mG^{SC}_{\dta}(a,\tau)
/\left[ 1+\mG^{SC}_{\dta}(a,\tau)\right]. & & 
\label{eq:G_SC_theta_def}
\end{eqnarray}

After having introduced the above two functions and after having
written down their explicit expressions, the generating function
$\mG\v(\tau)$ in equation (\ref{eq:gener_f}) can be obtained through
the solution of the system of equations (Bernardeau 1994b),
\begin{eqnarray}
\mG\v(f(\Omega,\Lambda)\tau) &=& \mG^{SC}\v\left[f(\Omega,\Lambda)\tau\ 
{\sigma\left([1+\mG_{\dta}(\tau)]^{1/3}R_0\right)
\over \sigma(R_0)}\right];  \nonumber \\
\mG_{\dta}(\tau) &=& \mG^{SC}_{\dta}\left[
\tau\ { \sigma\left([1+\mG_{\dta}(\tau)]^{1/3}R_0\right)
\over \sigma(R_0)} \right].
\end{eqnarray}

In principle, the above system of equations can be fully solved 
to get full-fledged expressions for the coefficients $T_p$, like e.g. 
the ones for $T_3$ and $T_4$ 
\begin{eqnarray}
T_3(\Omega=1,\Lambda=0)\,&=& -{ 26 \over 7}- \gamma_1\,, \label{eq:def_T3general}\\
T_4(\Omega=1,\Lambda=0)\,&=&  {12088 \over 441}+{ 338\over 21}\gamma_1+
{7 \over  3}\gamma_1^2+{ 2 \over  3} \gamma_2\,,\label{eq:def_T4general}
\end{eqnarray}
where the $\gamma_p$ are the successive logarithmic 
derivatives of the variance with the smoothing scale, 
\begin{equation}
	\gamma_p\equiv{\d^p\log \sigma^2_{\theta}(R) 
	\over \d\log^p R}\,.
\end{equation}
For practical purposes, however, it is preferable to define a
simplified set of equations that forms a reasonable approximation to
the original system. Several useful approximations can be applied. The
first one is to assume that the power spectrum of the density and
velocity fluctuations can be accurately described by a power law,
\begin{equation}
	P(k)\propto k^n\,,\ \ \sigma^2(R)\propto R^{-(n+3)},
\end{equation}
leading to $\gamma_1 = -(n+3)$ and $\gamma_2=0$ in
Eqs.~(\ref{eq:def_T3general}) and (\ref{eq:def_T4general}).
Although this may not be exact in general, it may be justified by 
the fact that we saw previously that the corrections in the values of 
$T_3$ and $T_4$ induced by a variation of the index $n$ 
enter only weakly. 

A second approximation exploits the fact that the $\Omega$ dependence
of $\mG^{SC}_{\dta}(\tau)$ is extremely weak (see Bouchet et al. 1992,
Bernardeau 1992). This implies that equation~(\ref{eq:G_SC_theta_def})
reduces to
\begin{equation}
	\mG^{SC}\v[f(\Omega,\Lambda)\tau]\approx-{\Omega^{0.6}
	\tau{ \d\over \d \tau}\mG^{SC}_{\dta}(\tau)/
	\left[1+\mG^{SC}_{\dta}(\tau)\right]}\,.
	\label{eq:G_SC_theta_approx}
\end{equation}
Moreover, it is therefore also reasonable to approximate the 
function $\mG^{SC}_{\dta}(\tau)$, independent of the value of 
$\Omega$ involved, by the simple expression for $\mG^{SC}_{\dta}(\tau)$ 
for the case $\Omega=0$, 
\begin{equation}
	\mG^{SC}_{\dta}(\tau)\approx\left(1+{ 2\tau\over
 	3}\right)^{-3/2}-1\,,
	\label{eq:G_SC_delta_fit}
\end{equation}
so that equation~(\ref{eq:G_SC_theta_def}), via
equation~(\ref{eq:G_SC_theta_approx}), lead to the approximate
relationship
\begin{equation}
	\mG^{SC}\v(\tau)\approx{\tau\left(1+{\displaystyle 2\tau\over
	\displaystyle 3\Omega^{0.6}}\right)^{-1}}.
	\label{eq:G_SC_theta_fit}
\end{equation}
Using the above approximate expressions in solving the system of
equations in (\ref{eq:phi_G_relat}) then yields an approximate
expression for the relation between the spherical collapse density
contrast $\tau$ and the functions $\mG^{SC}_{\dta}(\tau)$ and
$\mG^{SC}\v(\tau)$,
\begin{equation}
	\tau\approx{\displaystyle 3\over \displaystyle 2} 
	\left(1+\mG_{\dta}\right)^{(n+3)/6}
	\left[\left(1+\mG_{\dta}\right)^{-2/3}-1\right],
	\label{eq:tau_G_delta_relat}
\end{equation}
and
\begin{equation}
	\tau\approx{3\over 2}\left(1-{\displaystyle 2\mG\v\over
	\displaystyle 3 f(\Omega)}\right)^{(n+3)/6}\ 
	\left[\left(1-{\displaystyle 2\mG\v\over \displaystyle 3
	f(\Omega)}\right)^{-1}-1\right].
	\label{eq:tau_G_theta_relat}
\end{equation}
While reasonable, these approximations turn out not to be as good for
the velocity field as for the density field. For example, using the
approximations (\ref{eq:G_SC_delta_fit}) and (\ref{eq:G_SC_theta_fit})
one would obtain $S_3\equiv\mg\delta^3\md/\mg\delta^2\md^2=5-(n+3)$
instead of $S_3=34/7-(n+3)$, and $T_3=-4+(n+3)$ instead of
$T_3=-26/7+(n+3)$.  This means that for example for $n=-1$ the
relative errors in $S_3$ and $T_3$ are in the order of $5\%$ and
$15\%$.

While the above expose explains the strategy for calculating the
coefficients $T_p$ to any order, the major purpose of calculating the
generating function~(\ref{eq:gener_f}) is to construct the PDF of the
local velocity divergence. This is achieved through a Laplace inverse
transform (Balian \& Schaeffer 1989),
\begin{equation}
	p(\theta)\d\theta=\int_{-\ii\infty}^{+\ii\infty}
	{\d y\over 2\pi\ii\sigma\v^2}\exp\left[-{ \varphi\v(y)\over
	\sigma\v^2}
	+{ y\theta\over \sigma\v^2}\right] \d\theta,
\end{equation}
where $\sigma_{\theta}$ is the variance of the distribution. The 
saddle point approximation of this integral 
\begin{eqnarray}
	p_{\theta}(\theta)\d\theta&=&-{1\over \mG\v'(\tau)}
	\left[{1-\tau\mG\v^{''}
	(\tau)/\mG\v'(\tau) \over  2\pi \sigma\v^2 }\right]^{1/2}
	\nonumber \\
	& & \times\exp\left(-{ \tau^2\over 
	2\sigma\v^2}\right)
	\d\theta,\ \ \ \ \mG\v(\tau)=\theta\,,
	\label{eq:PDF_approx}
\end{eqnarray}
yields an accurate prescription for the shape of PDF, in particular 
around its maximum (see Bernardeau 1994b for more details). 

By subsequently invoking the approximate
expressions~(\ref{eq:G_SC_delta_fit}) and (\ref{eq:G_SC_theta_fit}),
and using the fact that then equations~(\ref{eq:tau_G_delta_relat})
and (\ref{eq:tau_G_theta_relat}) can be inverted, one can find simple
analytic fits [Eq.~(\ref{eq:PDF_theor})] for the special case of
$n=-1$.

\section[]{$\theta$ from one velocity component: an illustrative example}

It is possible to estimate the bias $\mu$ and the variance of the
noise $\sigma_e$ involved in our calculation of $\theta$
[Eq.~(\ref{eq:theta_w_estim}, \ref{eq:theta_estim_model})] with the
simple following model : let us assume that the difference of velocity
between two given neighboring points is collinear to their relative
position $\Delta{\bmit r}$
\begin{equation}
	\Delta{\bmit v} = \Theta \frac{\Delta{\bmit r}}{\Delta r},
	\label{eq:model_trivial}
\end{equation}
leading to $\theta_{\rm wall} = \Theta$. This assumption is generally
not true, even for an irrotational
flow, because it overlooks any possible shear. However this should be
enough for a rough estimation of $\mu$ and $\sigma_e$, and to
show that they do not depend directly on $\Omega$.
Considering that only the velocity along the direction
$x$ is known, our estimate of the divergence
[Eq.~(\ref{eq:theta_w_estim})] yields
\begin{equation}
	\theta_{\rm wall}^{\rm estim.}=
	\Delta v_x { n_x\over n_x^2+\epsilon}=\Theta  
	{ n_x^2\over n_x^2+\epsilon}.
\end{equation}
If we now assume that the direction of $\Delta{\bmit r}$ is randomly
distributed in a 3 dimensional space, the bias is, according to our
empirical parametrisation~(\ref{eq:theta_estim_model}), (the subscript
``wall'' of $\theta$ will be dropped hereafter)
\begin{equation}
	\mu = \frac{\mg \theta_{\rm estim.}\, \theta \md}{\Theta^2} =
	\mg { n_x^2\over n_x^2+\epsilon} \md 
	= 1 - \sqrt{\epsilon}\arctan\left(\epsilon^{-1/2}\right).
\end{equation}
This implies that the noise 
\begin{equation}
	e = \Theta \left({ n_x^2\over \mu(n_x^2+\epsilon)}-1 \right)
\end{equation}
has a vanishing average, and its variance is
\begin{eqnarray}
	\sigma_e^2 & = &   \Theta^2 \left[
	{1\over\mu^2}\mg n_x^4\over\left(n_x^2+\epsilon\right)^2\md -1
	\right] \nonumber \\
	&=& {\Theta^2\over\mu^2}\left[1+{\epsilon\over{2(1+\epsilon)}}
	-{{3\sqrt{\epsilon}}\over2}\arctan\left({\epsilon}^{-1/2}\right)-
	\mu^2\right],
\end{eqnarray}
with $0\le\sigma_e^2\le 0.562\ \Theta^2$ for $0\le\epsilon\le 1$. 

We can now assume that $\Theta$ is itself a random variable {\em
independent} of $n_x$ such that $\mg\Theta\md = \mg\theta\md =0$ and
its variance $\mg\Theta^2\md = \sigma\v^2$ is given in
%HE Table~\ref{tab:cumulant}. This latest quantity does not have an 
    Table~\ref{tab:cumulant}. This variance does not have an 
explicit dependence on $\Omega$, and therefore $\sigma_e$ does not, 
while it does depend on the evolutionary stage of the system and the 
scale that is considered. On the other hand, $\mu$ is clearly
completely independent of $\Omega$. Analytical results for the bias 
and the noise
are given in Table~\ref{tab:par1D_anal} and can be compared to their
respective numerical values listed in Table~\ref{tab:par1D}.

Taking this model at face value would tell us to put $\epsilon=0$
giving $\mu=1$ and $e=1$, i.e. $\theta_{\rm estim.}$ would be a perfect
estimator of $\theta$. This is of course not true because of the
neglected shear component and of the vorticity likely to be present at
small scale, %HE <<
 not to mention the instability of such a numerical estimate. %HE >>
However,
in spite of its extreme crudeness, this model gives fairly good
analytical predictions of the `intrinsic' bias and noise introduced by
our numerical estimates of the velocity divergence, and indicates that
the remaining noise due to shear and vorticity is almost negligible.

\end{document}